# PCCP

## Accepted Manuscript

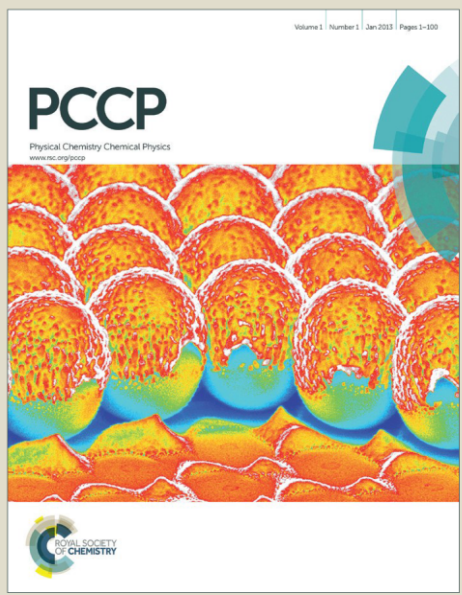

This is an *Accepted Manuscript*, which has been through the Royal Society of Chemistry peer review process and has been accepted for publication.

*Accepted Manuscripts* are published online shortly after acceptance, before technical editing, formatting and proof reading. Using this free service, authors can make their results available to the community, in citable form, before we publish the edited article. We will replace this *Accepted Manuscript* with the edited and formatted *Advance Article* as soon as it is available.

You can find more information about *Accepted Manuscripts* in the **Information for Authors**.

Please note that technical editing may introduce minor changes to the text and/or graphics, which may alter content. The journal's standard **Terms & Conditions** and the **Ethical guidelines** still apply. In no event shall the Royal Society of Chemistry be held responsible for any errors or omissions in this *Accepted Manuscript* or any consequences arising from the use of any information it contains.

# Magneto-thermally Activated Spin-state Transition in La$_{0.95}$Ca$_{0.05}$CoO$_3$: Magnetically-tunable Dipolar Glass and Giant Magneto-electricity

Suchita Pandey, Jitender Kumar[#], and A.M. Awasthi[*]

Thermodynamics Laboratory, UGC-DAE Consortium for Scientific Research,
University Campus, Khandwa Road, Indore- 452 001 (India)

## ABSTRACT

Magneto-dielectric spectroscopy of La$_{0.95}$Ca$_{0.05}$CoO$_3$ covering the crossover of spin states reveals strong coupling of its spin and dipolar degrees of freedom. Signature of spin-state transition at 30K clearly manifests in the magnetization data at 1Tesla optimal field. Our Co $L_{3,2}$-edge X-ray absorption spectrum on the doped-specimen is consistent with its suppressed low-to-intermediate spin-state transition temperature-- ~30K vis-à-vis ~150K documented on the pure LaCoO$_3$. Dispersive activation-step in dielectric constant $\Delta\varepsilon'_\omega(T) \sim O(10^2)$ with associated relaxation-peak in imaginary permittivity $\varepsilon''_\omega(T)$ characterize the allied influence of coexistent spin-states on the dielectric character. Dipolar relaxation in the low-spin regime below the transition temperature is partly segmental (Vogel-Fulcher-Tamman (VFT) kinetics) and features magnetic-field tunability, whereas in the low/intermediate-spin disordered state above, it is uncorrelated (Arrhenic kinetics) and almost impervious to the magnetic field $H$. Kinetics-switchover defines the dipolar-glass transition temperature $T_g(H)$ (=27K|$_{0T}$), below which their magneto-thermally-activated cooperative relaxations freeze-out by the VFT temperature $T_0(H)$ (=15K|$_{0T}$). Applied magnetic field facilitates thermal-activation in toggling the low-spins up into the intermediate-states. Consequently, the downsized dipolar-glass segments in the low-spin state and independent dipoles in the intermediate state exhibit accelerated dynamics. Critical 5Tesla field collapses the entire relaxation-kinetics onto a single Arrhenic behaviour, signaling that the dipolar-glass is completely devitrified under all higher fields. Magneto-electricity (ME) spanning sizeable thermo-spectral range registers diverse signatures here in the kinetic, spectral, and field behaviors, in contrast to the static/perturbative ME observed close to the spin-ordering in typical multiferroics. Intrinsic magneto-dielectricity (50%) along with vanishing magneto-loss is obtained at (27K/50kHz)$_{9T}$. Sub-linear deviant and field-hysteretic split seen in $\varepsilon'_{\omega,T}(H)$ above 4Tesla suggests the emergence of robust dipoles organized into nano-clusters, induced by the internally-generated high magneto-electric field. An elaborate $\omega$-$T$ multi-dispersions diagram maps the rich variety of phase/response patterns, revealing highly-interacting magnetic and electric moments in the system.

[*] Corresponding Author e-mail: amawasthi@csr.res.in. Tel: +91 731 2463913.
[#] Present Address: Indian Institute of Science Education and Research, Homi Bhabha Road, Pashan, Pune- 411 008 (India).

**Introduction**

Multiferroics (MF) as functional materials offer loss-free and fast cross-control of quad-state spintronics configurations via coupled electrical and magnetic order-parameters [1]. A most favorable situation is realized in the magneto-electric (ME) materials where two different ferroic states e.g., ferro/antiferroelectric and ferro/antiferromagnetic, coexist [2]. Besides the much desirable yet scarce linear ME coupling in the MF's featuring magnetic/electrical ordering (type-I&II multiferroics), there is ample scope for magneto-dielectricity (MD) in materials having quench-disordered or vitreous spin and/or polar degrees of freedom ("type-III" multiferroics or "multiglasses"). For example, in $Sr_{0.98}Mn_{0.02}TiO_3$ [3] and $K_{0.97}Mn_{0.03}TaO_3$ [4], appreciable MD coupling exists between their polar and spin glass attributes, $CuCr_{1/2}V_{1/2}O_2$ is magnetic and dipolar glass [5] with different freezing temperatures for magnetic and electrical moments, partially-disordered double-perovskite $La_2NiMnO_6$ [6] exhibits large room temperature MD (8-20%) across its wide polar glass regime, and MD comes from spin-reorientation in $La_{1/2}Sr_{1/2}NiO_4$ [7]. Substantial non-linear ME has been obtained in the antiferromagnetic dipole glass $CuCr_{1-x}In_xP_2S_6$ [8] and the spin-glass $PbFe_{1/2}Nb_{1/2}O_3$ [9]. In a recent study [10], magneto-electricity was found to manifest an emergent quantum paraelectric glass (QPG) state in $SrCu_3Ti_4O_{12}$.

Rare-earth $LaCoO_3$ (LCO) has been vigorously investigated [11] for its thermally-activated spin-state transition (SST). In the cobaltate, disorder-broadened crossover proceeds with increasing temperature from the low-spin ($^1A_1, t_{2g}^6 e_g^0$; $S$ =0) state through the intermediate-spin ($^3T_1, t_{2g}^5 e_g^1$; $S$ =1) to the high-spin ($^5T_2, t_{2g}^4 e_g^2$; $S$ =2) state [12]; LS→(LS-IS)→IS→(IS-HS) [13]. The $t_{2g}^5 e_g^1$ configuration gives higher dielectric constant probably due to a Jahn-Teller (JT) lattice distortion in the IS state; though recent local-distortion studies have challenged this as the cause in $LaCoO_3$ [13, 14]. Magnetic and other measurements determined the low-spin (LS) to intermediate-spin (IS) transition in $LaCoO_3$ around 150K [11]. SST also occurs in other cobaltates such as $Pr_{1/2}Ca_{1/2}CoO_3$ [15] and $GdBaCo_2O_5$ [16]. LCO exhibits trigonal symmetry with space group $R\bar{3}c$, and small monoclinic distortion due to orbital ordering [3]. Dispersive permittivity in $LaCoO_3$ [17] is understood as resulting from the nanoscale coexistence of monoclinic IS and rhombohedral LS phases [18]. The latter is evident in the bi-splitting of the La-O vibrational-mode degeneracy above 120K [17, 19], attributed to local symmetry-lowering that accompanies the LS→(LS-IS) transition. As with antiferroelectric & ferroelectric domains in relaxors, phase coexistence in cobaltates with Co-ions in different electronic structures is responsible for quantitative similarity of their dielectric dispersions.

Schmidt et. al. [20] reported magneto-dielectricity in the rare-earth cobaltates across magnetic spin-polaron defect region, which indicated the presence of spin-dipole coupling. Beside these, hole-doped (La-site) cobaltates show unusual behaviour vis-à-vis pure LCO. In $La_{1-x}A_xCoO_3$, divalent cation $A^{2+}$ (such as $Sr^{2+}$ and $Ca^{2+}$) generates extra charge state of Cobalt ($Co^{+4}$) with prominent effects; at lower concentrations of Sr the system is disordered and shows spin-glass behaviour (below $x = 0.18$). With further doping, the system becomes optimally ferromagnetic by $x = 0.5$ (equal concentrations of $Co^{+3}$ and $Co^{+4}$ cations), which facilitates double exchange in the system [21]. We have mildly replaced La in $LaCoO_3$ by (5%) Ca (hereafter called LCCO5); prospecting that with reduced SST temperature, dipolar (spin) interactions may realize quench-disordered/multiglass system, and exhibit appreciable magneto-dielectricity over wide thermo-spectral window. That Ca-doped $La_{0.95}Ca_{0.05}CoO_3$ undergoes the spin-state transition at a lower temperature vis-à-vis pure $LaCoO_3$ is confirmed by the magnetization results, supported via our Cobalt $L$-edge XAS spectra, and consequential to the electrical transport. Our magneto-dielectric measurements on LCCO5 reveal that the coupled electrical/spin state is altered upon Ca-doping, and huge intrinsic MD effects are obtained under the application of up to 9T field. Permittivity spectra are carefully analyzed to separate the low-frequency extrinsic/conductive (Maxwell-Wagner/free-charge) and high-frequency intrinsic/relaxational (dielectric-dipolar) responses. Genuine magneto-dielectricity is identified within intrinsic regime, uncorrelated to the magneto-conductive part, over particulate thermo-spectral band.

## Experimental

Ceramic $La_{0.95}Ca_{0.05}CoO_3$ sample was prepared by standard solid state reaction method using high-purity (99.99%) powders of $La_2O_3$, $CaCO_3$, and $Co_3O_4$. As most of the rare-earth oxides absorb atmospheric moisture, $La_2O_3$ was preheated at 1000°C for 6hrs. All the powders were mixed in stoichiometric ratio, ground and calcined at 1000°C and 1100°C for 24hrs, and then pressed into disc-shaped 2.5 mm thick pellets of diameter 10 mm. The pelletized samples were sintered at 1200°C for 20hrs and silver-coated for good electrical contacts for dielectric measurements. XRD characterization of the sample was done with Bruker D8 Advanced Diffractometer using Cu-$K_\alpha$ radiation. It was found that the desired compound formed in single phase with no detectable secondary phase, and the diffraction pattern could be indexed purely with the trigonal space group $R\bar{3}c$ perovskite-like crystal structure. Magnetization measurements were done on VSM-SQUID (Quantum Design) up to 7 Tesla. Room temperature soft X-ray absorption spectroscopy (XAS) was performed across Co $L_{3,2}$ edges in LCCO5 at beamline BL-01, Indus-2 (RRCAT, Indore Synchrotron Source) in total electron yield (TEY) mode. Base-pressure in the XAS chamber during the measurements was ~ $10^{-10}$ mbar and energy-resolution at Cobalt $L$-edge energy was ~ 300 meV. Broadband permittivity spectra were collected with 1V ac-excitation, using NOVO-CONTROL (Alpha-A) High Performance Frequency Analyzer. Magneto-dielectric characterization (up to 9 Tesla) was done using an Oxford NanoScience Integra System Magnet-Cryostat along with the Novo-Control Analyzer.

## Results and Discussion

Figure 1 shows the room temperature XRD pattern of our $La_{0.95}Ca_{0.05}CoO_3$ sample, collected using Cu-$K_\alpha$ radiation ($\lambda = 1.5405$Å). Within the sensitivity limit of Brucker D8 XRD set up, no extra peaks have been observed. All the observed reflections can be very well indexed within the trigonal space group $R\bar{3}c$, similar to the parent LaCoO$_3$. In order to accurately determine the lattice parameters as well as to confirm the designated space group, XRD pattern has been fitted using LeBail peak profile program (Full-Prof software, profile matching with constant scale factor) and the goodness of fit ($\chi^2$) was 1.52. The lattice parameters are obtained as: $a$, $b = 5.4430(1)$Å, and $c = 13.0993(3)$Å.

Magnetization versus temperature and field is shown in figs.2 and 3 respectively. Splitting of zero-field cooled warming (ZFCW) and field-cooled warming (FCW) $M(T)$ data (fig.2a insets) indicates some magnetic-ordering at ~90K. In addition, clear slope-discontinuity (with shallow notch) at 30K (ZFCW) and ~29K (FCW) is well discerned in the $M$-$T$ data at 1T. This anomaly signifying the spin state transition (SST) vanishes at higher (7T) and lower fields, and is completely washed out at 100 Oe (lower inset, fig.2a). The particulate SST signature can be understood as follows. The low-spin (LS) to intermediate-spin (IS) transition on warming involves increase in both the magnetic moment ($\mu$) and the energy-split ($\Delta \sim \mu H$) between its up/down configurations under the applied field. As regards the sample magnetization $M(T)$, these increases mutually compete/oppose across $T_{\text{SST}}$; increased-$\Delta$ wins first to *slightly reduce* $M(T_{\text{SST}}^-)$ then increased $\mu$ *slightly raises* $M(T_{\text{SST}}^+)$, versus an otherwise-smoother nominal $M(T_{\text{SST}})$. Finally, increased $\mu$ gives a *higher* and increased $\Delta$ gives a *less-steeper* $M(T > T_{\text{SST}})$ baseline, vis-à-vis that extrapolated from $M(T < T_{\text{SST}})$. Figure 2b shows the discontinuous slopes ($dM/dT$) in ZFCW and FCW $M(T)$ data taken at 1Tesla field, with appreciable jumps at $T_{\text{SST}}$ and $\left|dM/dT\right|_{T_{\text{SST}}^-} < \left|dM/dT\right|_{T_{\text{SST}}^+}$ as noted above. Evidently, considerations of relative changes in $\mu$, $\Delta$, and $M$ across $T_{\text{SST}}$ determine the observability here of the sharp $T_{\text{SST}}$-anomaly at 1Tesla optimal field. Note that the $T_{\text{SST}}$-anomaly is more pronounced in the field-cooled (FCW) data.

Figure 3 shows the $M(H)$ data taken at 25, 30, and 35K. At 25K, the low-field/linear regime is presumably due to the LS-rich state; the local-susceptibility ($dM/dH$) points to the onset of nonlinearity at approx. 400 Oe and eventually displays power-law behaviour in $H$ from 4T onwards (fig.3 lower-right panel). The mostly IS-rich state at 35K (having larger $\mu$ and $\Delta$) understandably enters non-linearity at a higher field (~ 0.1T), with ($dM/dH$) following power-law in $H$ beyond 3Tesla. A wider crossover field-span (4T/400 Oe =100) for 25K data (*despite* lower temperature) versus that for 35K (3T/0.1T=30) evidences LS→IS conversions at 25K stretching the otherwise Brillouin field-window. Convergence of 25 & 30K local-susceptibility with that at 35K beyond 1.5Tesla (upper-left/lower-right panels) signals the establishment of robust IS state, complementing the $T_{\text{SST}}$-anomaly seen at 30K in $M(T)|_{1T}$, besides explaining its absence in $M(T)$ at lower and higher fields.

Soft X-ray absorption spectroscopy is an ideal tool to investigate the valence states and phenomena like spin state transitions (SST) by measuring the strength of core shell electrons' transition to the unoccupied states (absorption phenomenon). XAS studies at Cobalt $L_{3,2}$-edges probe the vacancies in partially-filled $3d$ levels via the electronic transitions into them from spin-orbit split $2p_{3/2}$ and $2p_{1/2}$ levels, termed as $L_3$ and $L_2$ edges, respectively. Here, relative populations in $t_{2g}$ and $e_g$ sub-bands of $3d$ levels varies with the Co spin-state (LS: $t_{2g}^6 e_g^0$, $S=0$; IS: $t_{2g}^5 e_g^1$, $S=1$; HS: $t_{2g}^4 e_g^2$, $S=2$) and consequently, peak-profiles of $L$-edges reflect its variation. It is to be mentioned that in several of hole-doping (as large as up to 30%) studies in LaCoO₃, absence of edge-shift signatures rules out $Co^{3+}$ valence states' variation [22]; so our analysis of peak-profile is focused to highlight/estimate the spin states in LCCO5. In fig.4 we have plotted our room temperature XAS at Cobalt $L$-edges in LCCO5 along with the same in pure LaCoO₃ by Haverkort et. al. [23], by matching their white-line peaks. Excellent agreement in pre-edge and post-edge lines of these normalized spectra clearly shows that a one-to-one correlation of the features in these spectra can be made, which is not resolution-limited. Features marked with arrows in our spectra represent the broadening/extra-intensity, reflecting increased vacancy in $t_{2g}$ sub-bands, and consequently a higher fraction of the HS-state at room temperature in the Ca-doped vs. the pure specimen. Further, our room temperature XAS resembles much more with that of the pure LaCoO₃ at 650K [23]. Notably, under-the-peaks area-differences are found as ~10.7% (LCCO5/300K vs. LCO/300K), ~8.9% (LCO/650K vs. LCO/300K), and ~1.65% (LCCO5/300K vs. LCO/650K). From their temperature-dependent XAS, Haverkort et. al. determined the HS-state population-increase from 25% at 300K to 40% at 650K in LaCoO₃; most probably, HS-occupation in our LCCO5 is about ~40% at room temperature itself. Therefore vs. LaCoO₃, higher HS-state-fraction at 300K indicates that the transition into the LS-state is expected to occur in LCCO5 at a *lower* temperature than in LaCoO₃ (~150K, SST from dielectric), as confirmed by our magnetization data, having implications for the magneto-electric transport.

Figure 5 shows $\varepsilon''(\omega)$ at 20K and 40K, fitted (WinFit Software/NovoControl) using the Havriliak-Negami (H-N) equation [24]:

$$\varepsilon*(\omega) = \varepsilon' - i\varepsilon'' = \frac{\Delta\varepsilon}{\left[1+\left(i\omega\tau\right)^{\alpha}\right]^{\beta}} - i\left(\frac{\sigma_0}{\varepsilon_0\omega}\right)^n + \varepsilon_\infty \qquad (1)$$

This equation accurately estimates the relaxation time and shape-parameters of the relaxation-peak. Here, $\omega=2\pi f$ is the probing (angular) frequency, $\varepsilon_0$ denotes the vacuum permittivity, $\tau$ is the mean relaxation time, $\alpha$ and $\beta$ (both 0 to 1) are the spectral -width (broadening) and -shape (asymmetry) parameters of the relaxation peak $\varepsilon''(\omega_p)$, and $\Delta\varepsilon = \varepsilon'_0-\varepsilon'_\infty$ is known as the *dielectric strength* of the material. Fig.5 shows that the lower frequency region dominated by the Maxwell-Wagner effect is confined below 7Hz/0T (15Hz/9T) at low temperature (20K), while at high temperature (40K) it extends up to 8kHz/0T (10kHz/9T). Applied 9T field accelerates dipolar relaxations in the mixed (LS-IS) state at 20K much more (by +4.6dB) than those in the robust IS state at 40K (only by +0.84dB), of crucial consequence to ME transport. Cole-Cole plot (inset) of $\varepsilon''$ vs. $\varepsilon'$ at 40K/0T clearly delineates the well-defined extrinsic (conductance-dominant/M-W, $\varepsilon' \propto \varepsilon'' \equiv \varepsilon_0\sigma'/\omega$) and intrinsic (relaxational, semicircle) regimes; the intrinsic response in LCCO5 turns dramatically broadband only below ~30K.

Temperature dependent dielectric constant at selected frequencies, measured under zero and 9T magnetic field is shown in fig.6a. It gives giant dielectric constant over a wide $T$-range with room temperature value ~3×10⁴ at 1kHz (upper inset), which is much larger than that reported in the Mn-doped LaCoO₃ [25]. Observation of dielectric steps around ~150K across the spin states in pure

LaCoO$_3$ is already reported by Sudheendra et. al. [17]; with Ca-doping at La-site, a broadened dielectric step with magnitude ~$O(10^2)$ appears here at lower temperature ~30K. Broadened step is a characteristic signature of disorder, i.e., the coexistence of different spin states, and is resolved here into three regimes. (1) In the lower temperature regime (≤10K), robust low-spin (LS) state exists with little effect under the applied $H$-field. (2) Step-rise region (around 30K) where intermediate spin (IS) islands nucleate in the LS matrix [17] (LS-IS coexistent region). The longer Co-O bond due to larger J-T distortion in the IS state corresponds to higher dielectric constant with temperature-rise, resulting from more admixture of the IS-states. This region offers optimum scope for large magneto-dielectric effect, due to magnetically-tunable spin-state admixture. It is to be mentioned that this type of collective behaviour is characteristic in the spin-reorientation type materials, such as La$_{3/2}$Sr$_{1/2}$NiO$_{4+\delta}$ [7]. (3) Higher-temperature IS-rich regime, with relatively field-impervious single-particle dipolar-relaxation again exhibits much mitigated magneto-dielectricity.

Step-like increase of the dielectric constant ($\varepsilon'$) is associated with a relaxation peak in the imaginary part of permittivity $\varepsilon''$ (shown in fig.6b, the higher-$T$ data for each frequency have been clipped for clarity). Relaxation peaks in $\varepsilon''$ vs. $T$ shift towards the higher temperature with increasing probe frequency, reflecting the activated character of relaxations. The observed relaxation behaviour is different from that of colossal dielectric constant (CDC) materials [26], and more akin to the activation associated with spin-reorientation, as has been observed in La$_{1.5}$Sr$_{0.5}$NiO$_{4+\delta}$ [7]. Application of magnetic field is seen to steepen the $\varepsilon'$-step of the dielectric permittivity and cause $T$-downshift ~$O(1K)$ in the $\varepsilon''$-peak temperatures (fig.6b), within the mixed (LS-IS) regime. This is due to the larger dielectric constant of the field-induced/extra IS states and accelerated dynamics of the independent dipoles associated with the same. Further, with field-facilitation of the LS→(LS-IS) thermally-activated transition, size-reduced cooperative-dipoles (LS) undergo less-sluggish relaxation. An important implication of the *zero-field* data in fig.6b is the variation of relaxation-kinetics character across its broad spin-state crossover temperature, as made precise in the following.

We have analyzed the relaxation-kinetics viz., thermal-dispersion of spectral peak position in imaginary permittivity $\varepsilon''_T(\omega)$, with Arrhenic and/or Vogel-Fulcher-Tamman (VFT) [27] behaviors, respectively given below

$$\tau = \tau_{0A} \, \text{Exp}(E_a/k_B T), \tag{2}$$

$$\tau = \tau_{0V} \, \text{Exp}[U_a/(T-T_0)] \tag{3}$$

Here $\tau_0$ is a characteristic/approach time as $T \rightarrow \infty$, $E_a$ is activation energy for single-dipole relaxations, $T_0$ is the VFT temperature (freezing temperature) and $U_a$ is the (apparent) activation energy, of correlated-dipoles' relaxation process [27]. Arrhenius plots of the effective relaxation time $\tau_p(T) = 1/\omega_p$ against the corresponding inverse temperature $10^3/T$ are shown in fig.6a lower-inset. The low-temperature part at zero-field provides the glassy/VFT-parameters as $U_a(0T) = 2.21$meV, $T_0(0T) = 15$K, $\tau_{0V}(0T) = 1.47 \times 10^{-5}$ sec and the high-temperature part exhibiting the linear/Arrhenic kinetics has activation energy $E_a(0T)= 26\text{meV}$ and $\tau_{0V}(0T)= 6.71\times10^{-10}$ sec. Orders of magnitude gap between $\tau_{0V}(0T)$ (in LS) and $\tau_{0A}(0T)$ (in IS) here relates to the widely different polaronic characters in the two spin states, responsible for their dipolar-responses [25, 28]. This huge increase of the approach-timescale over-compensates the observed *anomalous decrease* of barrier-activation energy upon vitrification (for an overall deceleration of relaxation-dynamics at lower temperatures); $U_a(0T) < E_a(0T)$, contrary to latter's *expected increase* [29] when the elemental dipoles' integrity remains intact across the glass transition.

Under high magnetic field (9T), relaxation kinetics becomes globally Arrhenic (lower inset, fig.6a), with the activation energy $U_a(9T) \equiv E_a(9T)$ ~24meV, $T_0(9T) =0K$, and $\tau_{0V}(9T) \equiv \tau_{0A}(9T)$ $=1.14\times10^{-9}$ sec, reflecting unsegmented-dipoles' relaxation. Thus, there is relatively little change in the high-$T$ activation-energy under the applied magnetic-field ($\varDelta E_a(9T)/E_a(0T)$~ 7%), also merged to at high-$H$ by the low-$T$ apparent activation-energy $U_a(9T)$. This evidences that thermal-activation/kinetics of *uncorrelated* dipoles is rather impervious to the spin-state (LS, IS, or HS) they are associated with; although as we shall witness later, their *spectral-behaviors* to the *applied* magnetic field may differ. Notice that $\tau_{0A}(0T) < \tau_{0A}(9T)$ already indicates some ME-field-induced organization of the dipolar-entities in the IS state, which, due to their higher inertia (larger-sizescale) eventually ($T\rightarrow\infty$) relax slower vis-à-vis their "bare"/zero-field counterparts' high-$T$ dynamics. Below $T_g$, field-activated (LS→LS-IS) switching of the character of relaxing entities from glassy/cooperative to independent, is thus also vindicated energetically and kinetically, as the dominant mechanism of magneto-electricity in the system. Here, the optimal scope for this crucial switching is seen to exist below $T_g(0T)= 27K$ down to ~15K (fig.6a). This circumstance solely emerges from the *lowered* spin-state transition-temperature in LCCO5 (resulting in *different* Arrhenic (VFT) kinetics above (below) roughly $T_{SST} \approx30K$) vis-à-vis pure LCO (featuring *entirely* Arrhenic kinetics above *and* below $T_{SST}$ $\approx150K$, [17]). This $T_{SST}$-lowering endows the doped-specimen with much higher propensity for (and the realization of) the dipolar-glass state (segmentation of dipolar-dynamics, unperturbed by thermal-fluctuations). Further, as per expected, the Arrhenic activation energies of the doped (25meV) and the pure specimens (120meV, [17]) scale directly with their respective transition temperatures, by comparing their eq.2 for same $1/\tau \sim 8.25\text{kHz} = f_p(30K)$ and the expected universal $\tau_{0A}$ for IS-state dipoles in both LCO & LCCO5.

The observed variation in the behavior of $\varepsilon''$ vs. $T$ under zero and finite magnetic field can be understood as follows. At zero-field, Arrhenic kinetics results from the independent-dipoles (purely Debyean, exponential relaxations) above ~30K while below, cooperative-dipoles relax nonlinearly (Kohlrausch-Williams-Watts, stretched-exponential relaxations [30]), as manifest in the VFT kinetics (lower inset of fig.6a). Over the VFT-region, divergence of relaxation time by the finite freezing temperature $T_0(0T)=15K$ evidences segmentation of the relaxing entities [31]. In the high-temperature regime, Arrhenic fitting indicates unsegmented dipole-relaxations, evidently little-dependent on the applied $H$-field. Change in the relaxation type at ~27K defines the glass transition temperature $T_g(0)$, and for $H \geq 5T$, Arrhenic behaviour (single-particle relaxation) is recovered throughout $\varepsilon''$-peak temperature range, signifying complete devitrification of the low-temperature dipolar-glass phase. It is important to note however, that the persistence of observed bi-dispersion over the broad thermo-spectral domain qualitatively distinguishes the dipoles' organization here from that in relaxor, ferroelectric, and even unordered paraelectric states. The glass transition temperature is also manifest in the thermal behavior of relaxation-broadening as follows. Inset of fig.6b shows the differential--"magneto-broadening" parameter obtained from relaxation peaks at 9T and 0T; $\Delta\alpha(T) = \alpha_{9T}-\alpha_{0T}$. Saturated values of $\alpha_{0T}$ and $\alpha_{9T}$ above ~30K give a very small and almost $T$-independent $\Delta\alpha(T)$, whereas at lower temperatures $\Delta\alpha(T)$ rises sharply. Abrupt change at the onset temperature of 26.8K essentially reflects the sudden drop in $\alpha_{0T}$ below the glass transition temperature (no vitreous state at 9T field), made steeper by taking the differential. The sluggish dipolar-dynamics slows down and inhomogenizes (spectrally-broadens) the relaxations much more rapidly below $T_g$ [32].

The precision kinetics and its systematic evolution with the applied $H$-field allow the evaluation of field-dependent VFT-parameters in the LS glass-phase, as it retreats to lower temperatures with the increasing field. In fig.7 are shown $T_0(H)$ (left $y$-axis), $U_a(H)$ (right $y$-axis), and $T_g(H)$ (right-inset), , evaluated from the VFT (Arrhenic) fits to the low (high)-$T$ dispersion-kinetics dataset, consistent with fig.6a lower-inset & fig.7 left-inset. VFT parameters at 5T (not explicitly shown, due to the $H$-axis broken for better representation) are almost the same as those at 9T. Glass transition temperature $T_g(H)$ (right inset) is readily identified from the VFT to Arrhenic kinetics-crossover at each $H$; it presents a novel ME effect on an unconventional  electrical characteristic-- other than permittivity and polarization. More precisely, with $T_0(H)$ and $U_a(H)$ implicit in eq.3, we realize that the LS→IS crossover is *magneto*-thermally activated, since the effects on $\tau$ (eq.3) and on $\varepsilon^*$ (fig.6) of increasing $H$ and/or $T$ are similar over here. With increasing field, monotonous drop of the freezing temperature $T_0(H)$ to absolute zero accompanies logarithmically-equanimous (order of magnitude) rise in the activation energy $U_a(H)$. This remarkably-large difference of the activation energy between glassy/non-vitreous phases in LS-state of LCCO5 is beside its ~5× reduction vs. that in the pure LCO [17], and partly compensates for the emergence of finite $T_0$ (rapid divergence of $\tau$ below $T_g$), for quasi-continuation of $\tau$ over the narrow VFT-Arrhenic crossover region. Intriguing *finite* initial field-derivatives of the VFT freezing/activation parameters ($dT_0/dH|_{H=0}$ and $dU_a/dH|_{H=0} \neq 0$) is noteworthy.

Practical aspects of $H$-field driven devitrification of the dipolar-glass phase are examined by the spectral behaviour of canonical magneto-capacitance/conductance/loss. The standard ME parameters, evaluated at the benchmark glass temperature $T_g(0T)=27K$ for the maximum 9T field used presently, are defined below and shown in fig.8a over six decades in frequency.

$$\text{MD} = [\{\varepsilon'(9T) - \varepsilon'(0T)\}/\varepsilon'(0T)] \tag{4}$$

$$\text{MS} = [\{\varepsilon''(9T) - \varepsilon''(0T)\}/\varepsilon''(0T)] = [\{\sigma'(9T) - \sigma'(0T)\}/\sigma'(0T)] \tag{5}$$

$$\text{ML} = [\{\tan\delta(9T) - \tan\delta(0T)\}/\tan\delta(0T)] \tag{6}$$

also their inter-relation, $\qquad$ $$\text{ML} = (\text{MS-MD})/(1+\text{MD}) \tag{7}$$

We find at the outset an all positive-definite magneto-capacitance (MD) peaking at 80% with a huge (24dB) dynamic range, even as the magneto-conductance/loss assume both ±ve values across the spectrum. The vertical lines mark special events as follows. (1) Signifies the crossover (at 400Hz/9T) from extrinsic (M-W) to intrinsic (dipolar) response. (2) Minimum magneto-conductance (at 1.2kHz). (3) Minimum magneto-loss (at ~4kHz), found coincident with the $\varepsilon''$-loss-peak maximum (and thus MS ≡0) at 5T field, signify total devitrification of the glass-phase in LS state. (4) Maximum magneto-capacitance (at ~13kHz), (5) Zero magneto-loss (ML=0, also equal magneto-capacitance/conductance MD=MS at ~50kHz), (6) Broad maximum magneto-conductance (at ~80kHz). (7) Equal magneto-capacitance/loss (MD=ML at ~200kHz). From genuine-multiferroic standpoint, one needs to isolate the regime featuring uncorrelated magneto-capacitance/conductance [33]. We identify the spectral-window between (5) and (7) having the desired characteristic; though MS is static here, dynamic MD alone is responsible for the *variation* in ML (viz., eq.7), with MD=50% and ML=0% at ~50kHz.

At $T_g$(0T)=27K, large magneto-dielectricity (LMD) is manifest via *magnetically-activated* LS→IS transitions over the ~6dB (50kHz -200kHz) band, due to the upper/lower-bounded lengthscale over which the applied field *optimally* disorders the otherwise ($H$=0) LS-state matrix into coexistent (LS-IS) state. The IS-islands created under the field (associated with higher $\varepsilon'$/+ve MD and faster/single-particle relaxation) by local "melting" of the LS state are thus typically *nano*-dispersed. This is verified by the ratio of the frequency-scales of the *interfacial* to *inter-grain* M-W contributions (100kHz:100Hz), in being *inverse-proportion* to their respective size-scales (nm:μm). The relative constancy (decrease) of MS (MD) over the intrinsic-ME band is understood as follows. Under applied $H$-field, the *increase* of $\sigma'$ (ac-conductivity) here is contributed to *only* by the relaxations of the dipoles *within* variously-sized "IS-droplets", whereas to that of $\varepsilon'$ (charge-storage propensity), another-- IS-LS *interfacial* (yet intrinsic) M-W mechanism contributes more (less) at lower (higher) band-side from bigger (smaller) interfaces [33]. Dispersively-shifted sets at key temperatures generate the conformal isotherms of MS and ML vs. MD (fig.8b & c), with the numbered-loci of special points corresponding to those in fig.8a. Over the sector bounded by line (5) and curve (7), local slopes $\Delta$MS/$\Delta$MD ~$O$(0) *and* $\Delta$ML/$\Delta$MD ~$O$(-1) really do affirm the regime as "intrinsically magneto-dielectric". A substantial LMD frequency-bandwidth (e.g., 11dB at 20K) is obtained only below ~$T_g$, in the LS-IS disordered spin-state regime. Table 1 compiles the maximum values and those where ML≡0 for MD(9T) at key temperatures. Relatively large MD along with its mild-variations vs. $T$ & log{$\omega_{m,0}(T)$} stand apart from its steep falls at 10K and 40K. The relevant $T$-window for intrinsic-LMD is therefore 15-35K, over which $\omega_{m,0}(T)$ too have smoother variations. At higher temperatures, the frequency-coverage of LMD shrinks rapidly (fig.8b, c), as the LS state diminishes, whereas at lower-$T$'s, freezing of dipoles in the robust LS-state matrix reduces the ME-effect substantially. The description for 9T also applies at lower-fields, though the scaled-down magnitudes shift the spectral-benchmarks variously.

Although the spectral behaviour of relevant magneto-parameters remains generically similar vs. the applied magnetic field, the explicit field-dependent behaviour reveals further novel physics regarding the character and organization of the dipoles. At the zero-field glass transition temperature 27K, magneto-capacitance (MD) and magneto-loss (ML) vs. $H$ for various frequencies (including those marked in fig.8) are shown in fig.9a&b respectively. Two main features here are the linearity of MD($H$) at low-fields and its sub-linearity with concurrent $H$(up/down) split at > 4Tesla fields. Also, the frequencies of benchmarked features (especially 1, 3, 4, 5, and 7 in fig.8) are roughly $H$-independent, since their corresponding isochrones maintain their mutual ordering versus the field, as mentioned earlier. Inset in fig.9a evidences the identity of results in ±$H$-field-scans, without any remnant/coercive effects persistent around $H = 0$ during the up/down/up protocol from +ve to zero to −ve & back to zero-field. Note that at the M-W/dipolar crossover MD(400Hz) stays low at < 5%, even as ML(400Hz) builds up to ∼ -15% at 9Tesla, rendering the lower frequencies rather undesirable from the ME-perspective. While it is reassuring to witness the integrity of the low-field (≤ 4T) linear-ME response maintained over the broad spectral bandwidth, the appreciable $H$(up/down) split and sub-linear MD($H$) above 4T-field imply the existence of "domains" and their "pinning". Definitely, the responding entities at high-fields are not exactly the same as the sole $E_{ac}$-induced dipoles at lower-$H$.

It is entirely plausible that the ME-induced large $E_{ME}$ electric-field at high-$H$ materializes robust dipoles and organizes them over short-range (resembling super-paraelectric clusters [34]), which get mutually-interlocked at increasing $E_{ME}$; thus the residual (hysteretic) permittivity effect which disappears by 4T (decreasing) field, below which the "clustering" is non-existent. We also note that the anomalous high-$H$ feature is well appreciable roughly over the intrinsic-LMD band (50kHz-200kHz), as determined from fig.8; topologically associating the high-field stabilized dipole-clusters with but the *optimal* IS-droplets created by the $H$-field. Therefore, under >4T field, while the relaxation pattern becomes spectrally uniform (single-Arrhenic), LMD over this intrinsic-band is dominantly-affected by the dipole nano-clusters in the optimal IS-droplets. This would *distinctly define* the LS-IS interface, responsible for the decreasing MD($\omega$), while keeping MS static, around ∼100kHz. Moreover, independent but yet-unclustered dipoles in the LS-IS state would largely be responsible for MD($\omega$) *outside* this band, at high-field. This naturally accounts for the intriguing features observed here, constituting yet another novel manifestation of the ME effect at high-fields. As per table 2, the direct permittivity-slope $\{(\varepsilon'_9-\varepsilon'_0)/\Delta H\}_{9T}$ of the dipolar nano-clusters, emergent at high-fields, regresses with the linear *magneto-dielectric coupling* (MDC ≡ $d\varepsilon'/dH$) at low-$H$ < 4T.

The rich diversity of ME features witnessed in $T$-, $\omega$-, and $H$-domains and rooted to the dipolar-entities associated with the disordered spin-states mandates an integrated presentation. Role of the controlling variables is systematized by the multi-dispersions shown in the diagram of fig.10, representing the relevant responses/entities covering 20-34K thermal window of ME-interest. The M-W/Dipolar (extrinsic/intrinsic) response crossover, as defined in fig.5 registers its dispersion/thermal-shift from ∼15Hz/20K to 3.4kHz/34K; shown for 9T field as setting the *effective* boundary, that for zero-field lying at frequencies ~ -2dB (-60%) *lower* on average. Subtle signature of LS-rich to coexistent (IS-LS) spin-state crossover is implicit in this curve itself, as a mild-inflexion around ~27K, below which the intrinsic-response boundary lays somewhat lower vis-à-vis it's extrapolation from that above. At frequencies above this *operational* lower-boundary and thus entirely within the intrinsic-regime onwards, first the minimum magneto-conductance ($MS_{min}$) is dispersed over 50Hz/20K to 9kHz/34K. Relaxation peaks under zero-field next manifest their kinetic-dispersions; Vogel-Fulcher (low-*T*'s) and Arrhenic (high-*T*'s), with their discontinuous crossover defining the glass transition temperature $T_g(0T) = 27K$. Locus of these crossovers vs. *H* traces the dispersion $\omega_g(T_g)$ of the field-dependent dipolar-glass transition temperature $T_g(H)$, finally merging at $T_g(\geq 4T)$ with the pan-Arrhenic dispersion of relaxation-kinetics onset above 4T-field. This onset dispersed over 100Hz/20K to 34kHz/34K at the *critical devitrification field* $H_{cd}$ = 5T concurs minimum magneto-loss ($ML_{min}$) and zero magneto-conductance ($MS\approx0$). Maximum magneto-capacitance, apparently related to the length-scale of most of the LS→IS-droplets created by the field, is next seen to be dispersed over ~450Hz/20K ($MD_{max}$=120%) to ~200kHz/34K ($MD_{max}$=60%). Typically, the dispersion curve for maximum *dielectric coupling* ($MDC \equiv d\varepsilon'/dH|_{H < 4T}$), coincident with $ML_{min}$ and $d\varepsilon''/dH|_0\approx0$, is found ~ -5dB below that for maximum MD. The landmark ML=0 (MS≡MD) dispersion curve (setting the lower-boundary for genuine magneto-dielectricity) extends over 1.5kHz/20K (MD at ~80%) to 760kHz/34K (MD at ~37%); this also serves to define a dimensionless measure of the *spin state disorder window* (SSDW=$\Delta T/T_{SST}$ ~1/2, table 1). Quite large MD combined with vanishing ML over this special curve underscore the highpoint of both the LCCO5 system-functionality and the effectiveness of the applied field. This dispersion signifies the onset of MS-independent MD and MD-driven ML, manifest as start of the quasi-static MS (fig.8a, b), whose broadly-humped maximum ($MS_{max}$) is next dispersed over 5.6kHz/20K to 950kHz/34K. Finally, the genuine MD's upper-boundary (at MD=ML, beyond which *dynamic* MS again starts affecting both MD and ML) is dispersed over ~20kHz/20K to 2.25MHz/34K, signaling the spectral-demise of ME-functionality. Thus, the functionally-suitable spectral-bandwidth/MD-dynamic-range (irrespective of the applied-field's magnitude) varies over (11dB/3.7dB)$_{20K}$ to (5dB/3.4dB)$_{34K}$. Primarily, over this band, high-field ($H > 4$T) $\boldsymbol{E}_{ME}$-stabilized dipolar nano-clusters (in IS-state) determine the electrical transport. It is emphasized here that sans the *H*-field axis, this essentially response-map lacks an exact one-to-one correspondence of its distinctly-marked regimes to specific/exclusive electro-magnetic phases of the compound. For example, the $\omega$-*T* region under $T_g(H)$ marked as "segmental-dipoles" ($0 < H < 5$T) for their distinctive contribution here, is also accessible to the "unsegmented" dipolar entities ($H > 5$T). Similarly, in the upper-band marked as "dipolar-clusters" ($H > 4$T), "unclustered-dipoles" ($H < 4$T) of the LS-IS state do contribute as well. Nonetheless, this multi-dispersions diagram may serve as a paradigm in identifying the thermo-spectral scope for genuine magneto-electricity, clear of the considerable and dynamic magneto-conductance effects to be reckoned with, as in the present case.

Finally, we place our findings in perspective against similar literature-reported studies. The LS-IS spin state transition investigated here in only *mildly-doped* $La_{0.95}Ca_{0.05}CoO_3$ uniquely features devitrification of a dipolar-glass phase, qualitatively different from the diffused-FM transition in the *half-doped* single crystal $LaMn_{1/2}Co_{1/2}O_3$, exhibiting only a switchover between independent-dipoles characters with different (Arrhenic) activations [25]. In case of SST in pure $LaCoO_3$, even the barrier-activation energies of independent dipoles in the LS and IS states are alike [17]; a scenario realized in LCCO5 here at high-fields. The most important *functional* consequence of the novel phenomenon in LCCO5 is a huge intrinsic MD (~80% at 20K/1.5kHz), and together with its pan-spectral broad linearity in the $H \leq 4T$ field manifests a giant and direct ME-coupling. This is in sharp contrast with the low (6%) MD reported below the disordered FM-$T_C$ in single crystal $LaMn_{1/2}Co_{1/2}O_3$ [25] and still-lower ($\leq 0.6\%$) MD reported in $CuCr_{1/2}V_{1/2}O_2$ multiglass [6]; latter being visibly quadratic in $H$ is due to the higher-order ME-effects. Yet another, recent study on the spin-cluster glass $Fe_2TiO_5$ [35] reported a miniscule ($\leq 0.04\%$) quadratic MD at vastly-high fields (~14T, negligible up to ~5T). Emergence of super-paraelectric-like nano-clusters at high (> 4T) fields is the other significant maiden signature of exceptionally strong ME coupling in LCCO5. Studies on polycrystalline/thin-film samples of e.g., $DyMn_{1/3}Fe_{2/3}O_3$ [36], $La_2MnNiO_6$ [6, 37], and $LaMn_{1/2}Co_{1/2}O_3$ [38] have also reported large dielectric constants and MD; though the low-frequency response (mainly by the M-W/inter-grain/electrode-boundary) focused on does not represent the high-frequency/intrinsic ME. The present study has carefully excluded these extrinsic effects, as well as identified specific high-frequency band where magneto-capacitance/loss are impervious of the magneto-conductance.

Topological disorder created by the mild-substitution of smaller/heterovalent-Ca at larger-La site in spin-active $LaCoO_3$ thus proves optimally effective for LMD as follows. (1) It raises the high-spin state population vis-à-vis that in the virgin cobaltate, simulating the heating effect and thus lowers the LS→IS transition temperature in the doped-specimen. (2) It creates highly inhomogeneous spin-state ensembles at low temperature, catalyzing the vitrification of the polaronic dipoles in the LS state. (3) The enhanced nanoscale LS-IS phase-coexistence (SSDW=$\Delta T/T_{SST}$ ~1/2 for LCCO5 (table 1) vs. ~3/8 for LCO [17], from a common 1.54kHz-990kHz intrinsic-bandwidth for their respective dispersions) further differentiates the polaronic characters in the LS and IS states dramatically, making *local lattice-distortions* [14] as more dominant in LCCO5, exceeding those due to J-T interactions. (4) Finally, devitrification of the dipolar-glass state allied with magneto-thermally activated LS→IS transition endows the LCCO5 cobaltate high-functionality/tunability with relatively modest external magnetic-field, featuring maidenly large and diverse magneto-electric effects. Otherwise direct (Co-site) substitutions reduce the Co-ions' *number* themselves, thereby contracting the scope & influence of LS→IS transitions, responsible for the peculiar ME characteristics witnessed here. For example, the single crystal $LaMn_{1/2}Co_{1/2}O_3$ undergoes only a diffused-FM ordering at ~150K [25], exhibiting mild ME at close by temperatures, sans any spin state physics.

## Conclusions

To summarize, the magnetic characterization of mildly-doped $La_{0.95}Ca_{0.05}CoO_3$ evidences its spin-state transition temperature suppressed to 30K vs. ~150K of the pure $LaCoO_3$. We have established the nature of the disordered transition in $La_{0.95}Ca_{0.05}CoO_3$ as magneto-thermally-activated; the accelerating-influence of magnetic-field on strongly allied dipolar degrees of freedom simulates temperature-increment effects. We witness diverse signatures of magneto-electricity here in thermal, spectral, kinetic, and field domains, for the first time. Magneto-dielectric study of the crossover reveals the low-temperature LS-rich state harboring magnetically tunable dipolar-glass phase, which is completely devitrified under the critical 5Tesla field. Significantly, the linear magneto-dielectric coupling ($d\varepsilon'/dH|_0$) is maximized and magneto-dielectric loss-rate vanishes ($d\varepsilon''/dH|_0=0$) exactly on $\omega$-$T$ locus of the relaxation-peak at 5T. Splitting of the LS-rich matrix into nanometric IS-droplets (LS→LS-IS disorder) under applied magnetic field is found to dramatically alter the organization of their associated dipoles; from glassy character in the LS-state to super-paraelectric-like (high-$H$) nano-clusters in the IS-state. This accounts well as the dominant novel mechanism for the large magneto-dielectricity (LMD, field-hysteretic and sub-linear above 4T) and covers a wide thermo-spectral window of spin-state disordered LS-IS phase-coexistence. The LMD is spectrally isolated from the extrinsic effects and also disentangled from the comparable magneto-conductance. It would be interesting to investigate in single crystals the implicit issue of local polarization of dipoles in the nano-clusters emergent at high-fields. Our study offers a template prescription to examine similar systems for unusual magneto-electricity.

## Acknowledgments

Dinesh Shukla is thankfully acknowledged for the XAS data on $La_{0.95}Ca_{0.05}CoO_3$ and its analysis. Authors sincerely thank Ram Janay Choudhary for providing the magnetization data and for the discussion of its analyzed results.

**Figure Captions**

**Fig.1.** Room-temperature X-ray diffraction pattern of $La_{0.95}Ca_{0.05}CoO_3$ powder sample, with peak profile fitting (LeBail program).

**Fig.2.** Temperature dependent magnetization $M(T)$ (top panel) and $dM/dT$ (bottom panel) displays slope-discontinuity, evidencing spin state transition at $T_{SST} \approx 30K$, clearly manifest in the optimal 1T data and completely smeared out at 100 Oe. Survey $M(T)$ shown in the top panel insets at 100 Oe and 1T indicate some magnetic ordering at ~ 90K (ZFCW/FCW data-split).

**Fig.3.** Field dependence of magnetization $M(H)$ and local susceptibility $dM/dH$ at 25, 30, and 35K. Bottom-right panel: the crossover field-span (bridging the initial/linear and the final/power-law regimes) is anomalously wider (4T/400 Oe = 100) at 25K vs. (3T/0.1T = 30) at 35K, traceable to the LS→IS transitions at 25K. High-field convergence of all three local susceptibility data denotes the materialization beyond 1.5T, of the robust IS-rich state at 25 and 30K; complementing the anomaly manifestation in optimal $M(T)|_{1T}$ at $T_{SST} \approx 30K$ (fig.2).

**Fig.4.** Measured Co $L_{3,2}$-edges XAS spectrum in doped $La_{0.95}Ca_{0.05}CoO_3$ (LCCO5) at room temperature and those reported in pure $LaCoO_3$ (LCO) at 300K and 650K [23]. Under-the-peaks aerial differences ~10.7% (LCCO5/300K vs. LCO/300K), ~1.65% (LCCO5/300K vs. LCO/650K), and ~8.9% (LCO/650K vs. LCO/300K) appropriate the room-temperature HS-state occupation in LCCO5 as high as 40% (as in LCO at 650K, being 25% at 300K [23]).

**Fig.5.** Dielectric spectra below (20K) and above (40K) the spin-state transition temperature (30K) illustrate the highly temperature-dependent range of the inter-grain/extrinsic Maxwell-Wagner contribution, excluded from intrinsic magneto-electricity considerations in the subsequent analysis. Note the much higher acceleration by applied 9T field, of the relaxations in the mixed (LS-IS) state at 20K (4.6dB) vs. that in the robust IS state at 40K (0.84dB), consequent for the giant ME effects. Inset clearly delineates the power-law (M-W/space-charge) regime at low frequencies and the intrinsic semicircular (dipolar-relaxation) response at higher spectral range.

**Fig.6.** Dielectric permittivity $\varepsilon^*(T)$ at several frequencies, over the spin-state crossover region. Huge $\varepsilon'$-steps (a) and associated relaxation peaks in $\varepsilon''$ (b) at zero-field and their observed shifts (accelerated dynamics) under 9T-field duly indicate the magneto-thermally-activated nature of the transition. Upper inset in (a) shows $\varepsilon'(T)|_{1kHz}$ up to room temperature. Lower inset (a): Arrhenius plot of the relaxation time $\tau(T)$ clearly delineates glassy (VFT/low-$T$) and single-particle (Arrhenic/high-$T$) regimes at 0T. Above 4T field, the entire kinetics collapses onto a single Arrhenic form. Inset (b): Magneto-broadening parameter ($\Delta\alpha = \alpha_{9T}-\alpha_{0T}$, eq.1) of the relaxation-peak in $\varepsilon_T''(\omega)$ marks the dipolar glass transition temperature $T_g(0T) \approx 27K$, matching with that determined from the VFT-Arrhenic crossover (inset a).

**Fig.7.** VFT analysis of the LS-state (low-$T$) relaxation kinetics (e.g., left inset) and its crossover to (LS-IS)-state (high-$T$) Arrhenic-behaviour furnishes $T_0(H)$, $U_a(H)$, and $T_g(H)$; magneto-characterizing the dipolar glass. Anomalous *increase* of barrier-activation energy upon devitrification [29] at high-fields is overcompensated by the huge ~$O(10^4)$ decrease of the approach time $\tau_{0V}(H)$. Pan-Arrhenic kinetics at 5T mutes $T_g$, as well as $T_0(5T)\to0$ and $U_a(5T)\to E_a(5T) \approx E_a(0T)$. Direct scaling of the Arrhenic activation energies of doped $La_{0.95}Ca_{0.05}CoO_3$ (25meV) and pure $LaCoO_3$ (120meV) [17], with their respective transition temperatures ($T_{SST}$ at 8.25kHz) of ~30K and ~150K, is reassuring. Moreover, the finite field-derivatives $dT_0/dH|_0$ and $dU_a/dH|_0$ are intriguing for their prospective implication.

**Fig.8.** Spectral behavior of magneto-dielectric quantities (eqs.4-6) at 9Tesla. Numbered lines mark the frequency (a) and the loci (b and c) of (1) MW-Dipolar crossover, (2) minimum magneto-conductance, (3) minimum magneto-loss (zero magneto-conductance), (4) maximum magneto-capacitance, (5) zero magneto-loss (equal magneto-cap./conductance), (6) maximum magneto-conductance, and (7) equal magneto-capacitance/loss. The genuine magneto-dielectric regime, where magneto-capacitance has no regression vs. magneto-conductance ($\Delta$MS/$\Delta$MD $\sim O(0)$ in b) and perfect anti-regression with magneto-loss ($\Delta$ML/$\Delta$MD $\sim O(-1)$ in c) is hatched ($\sim$6dB bandwidth in a) and sectored (b, c), between lines (5) & (7); it diminishes rapidly beyond 34K, with LS-IS coexistence yielding to IS-rich state. Arrows in (b, c) denote increasing frequency.

**Fig.9.** Magneto-capacitance (a) and magneto-loss (b) at $T_g$(0T) = 27K vs. applied magnetic field for the benchmark frequencies noted in fig.5. Linear magneto-capacitance at low-$H$ and its sub-linearity/hysteresis beyond 4Tesla evidence different dipolar-organizations as being responsible for the two behaviors. Consistent with the kinetics, thermal, and spectral features, while linear magneto-electricity stems largely from the un-segmented dipoles in the LS-IS mixed state, independent dipoles in the IS-state (stabilized & nano-clustered by the large induced $\boldsymbol{E}_{ME}$-field) exhibit the peculiar features over the genuine ME-bandwidth, as identified in fig.5. Inset: lack of zero-field remnance/coercivity confirms the $\boldsymbol{E}_{ac}$-induced nature of the dipoles in LS-IS state.

**Fig.10.** Multi-dispersions diagram across 20K to 34K and 10Hz to $\sim$MHz thermo-spectral domain represents the broad scope of all the relevant magneto-thermally-activated ME features. Complete devitrification of the dipolar glass phase in the LS state is represented by the relaxation-dispersion curve at critical 5T field; featuring the locus of maximum linear magneto-dielectric coupling ($d\varepsilon'/dH|_0$) along with vanishing magneto-dielectric loss-rate ($d\varepsilon''/dH|_0 \equiv 0$). Functionally-relevant ME regime is highlighted as the stripe spread across $\sim$10kHz to $\sim$MHz, covering frequency-bandwidth and dynamic-range of magneto-capacitance (11dB & 3.7dB)$_{20K}$ to (5dB & 3.4dB)$_{34K}$. Intrinsic ME frequency-range at some $\geq$13dB above that of the relaxation-peak dispersion characterizes the functional $\omega/T$-window with relatively lower dielectric-losses. See text for a comprehensive description and association of dominant electro-magnetic material-phases to the local $\omega$-$T$ regions.

**Table 1.** Compilation at key temperatures of the maximum MD and of that accompanying ML ≡ 0, versus their respective frequencies. Smoother variations of these special MD's vs. $T$ and $\log[\omega_{m,0}(T)]$, and of $\omega_{m,0}(T)$ delineate 15-35K as the relevant thermal range for genuine-ME study.

| $T$ (K) | 10 | 15 | 20 | 25 | 30 | 35 | 40 |
|---|---|---|---|---|---|---|---|
| $\omega_m(T)$ (Hz) | 1.4 | 8 | 440 | 5.5k | 45k | 250k | 500k |
| $MD_m$ (%) | 47 | 162 | 117 | 89 | 70 | 57 | 38 |
| $\omega_0(T)$ (Hz) | 5 | 31 | 1.54k | 21k | 173k | 990k | 2.13M |
| $MD_0$ (%) | 34 | 111 | 79 | 57 | 43 | 34 | 21 |

**Table 2.** At $T_g(0T) = 27K$, a large (~21dB) spectral variation (over 4kHz-1MHz) in the low-field linear magneto-dielectric coupling MDC $= (d\varepsilon'/dH)_0$ is obtained, to which the direct permittivity-slope at 9Tesla exactly correlates as $(\Delta\varepsilon'/\Delta H)_9 = 0.88[(d\varepsilon'/dH)_0]^{0.97}$. MDC-maximum near 4kHz (generally at ~ -5dB lower w.r.t. MD-maximum frequency for all $T$'s) and vanishing magneto-dielectric loss-rate $(d\varepsilon''/dH=0\approx MS$, fig.5a,b) track the relaxation-peak at critical 5T field (setting complete devitrification of dipolar glass, fig.7). At lower temperatures, MDC rises further; e.g. at 20K/120Hz, max. $(d\varepsilon'/dH)_0 \approx 263/T$ and $(\Delta\varepsilon'/\Delta H)_9 \approx 214/T$.

| $f$ (Hz) | 100 | 400 | $10^3$ | $4\times10^3$ | $1.4\times10^4$ | $2.5\times10^4$ | $5\times10^4$ | $10^5$ | $2\times10^5$ | $4\times10^5$ | $10^6$ |
|---|---|---|---|---|---|---|---|---|---|---|---|
| $(d\varepsilon'/dH)_0$ | 47.89 | 60.07 | 114.4 | 225.1 | 86.72 | 37.15 | 16.08 | 6.908 | 3.514 | 2.337 | 1.788 |
| $(\Delta\varepsilon'/\Delta H)_9$ | 37.27 | 47.34 | 79.51 | 193.0 | 74.46 | 30.57 | 12.63 | 5.783 | 2.994 | 1.764 | 1.293 |

**Fig.1**

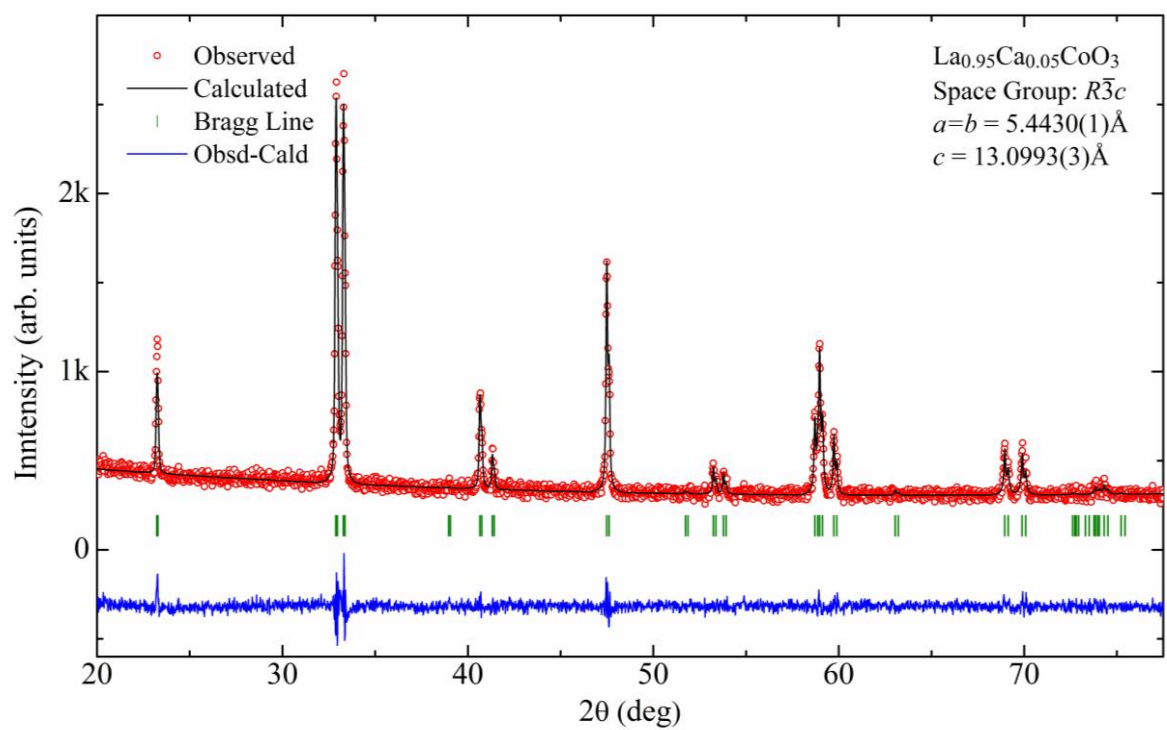

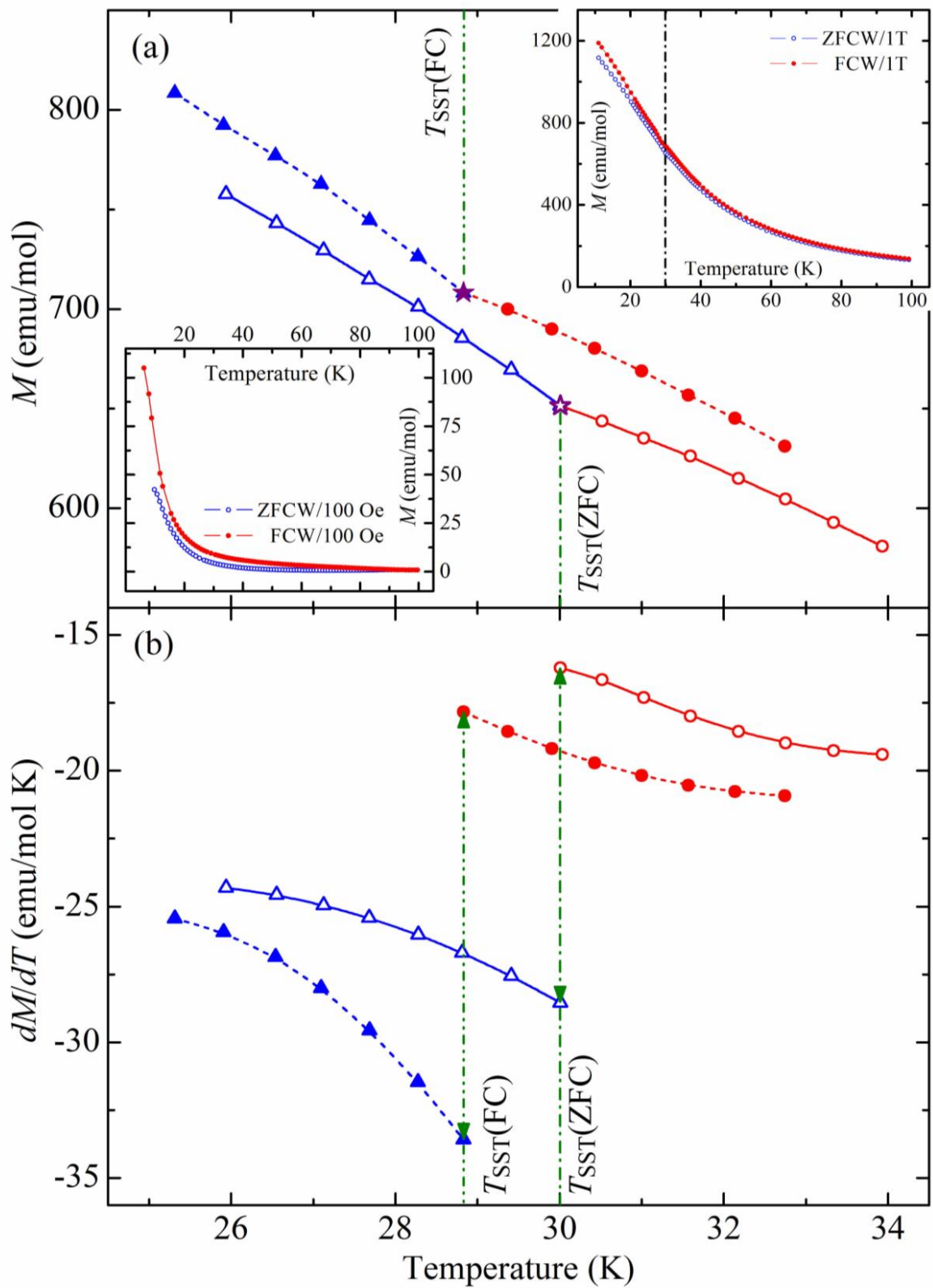

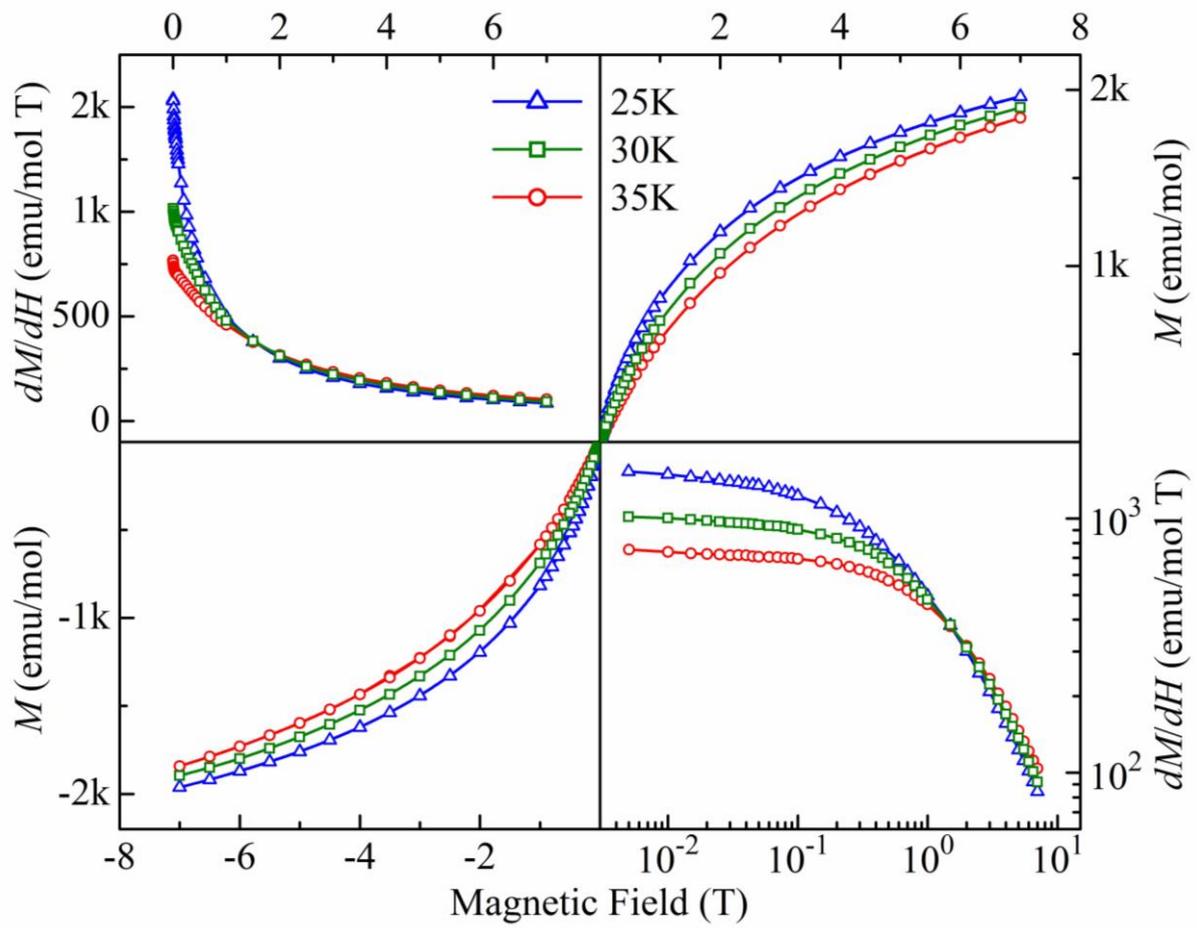

**Fig.4**

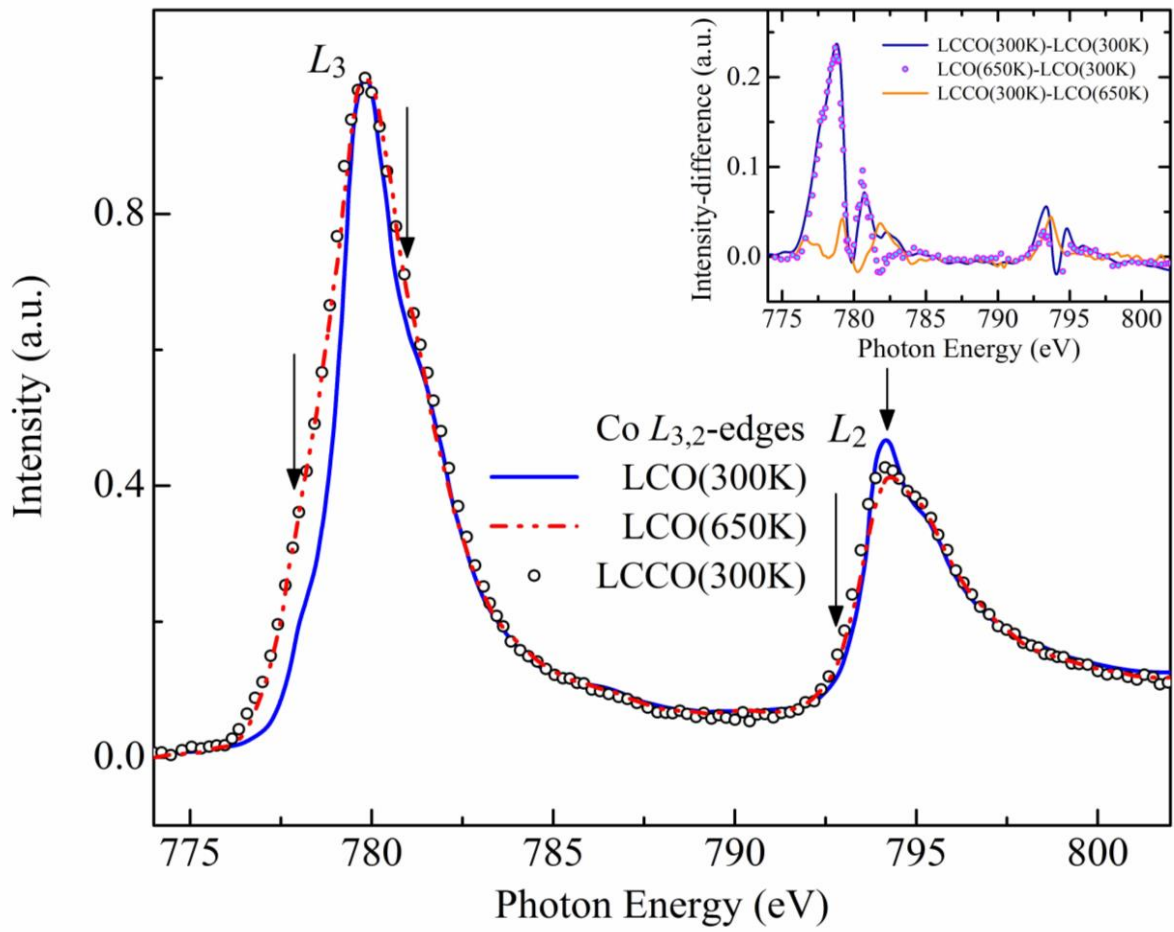

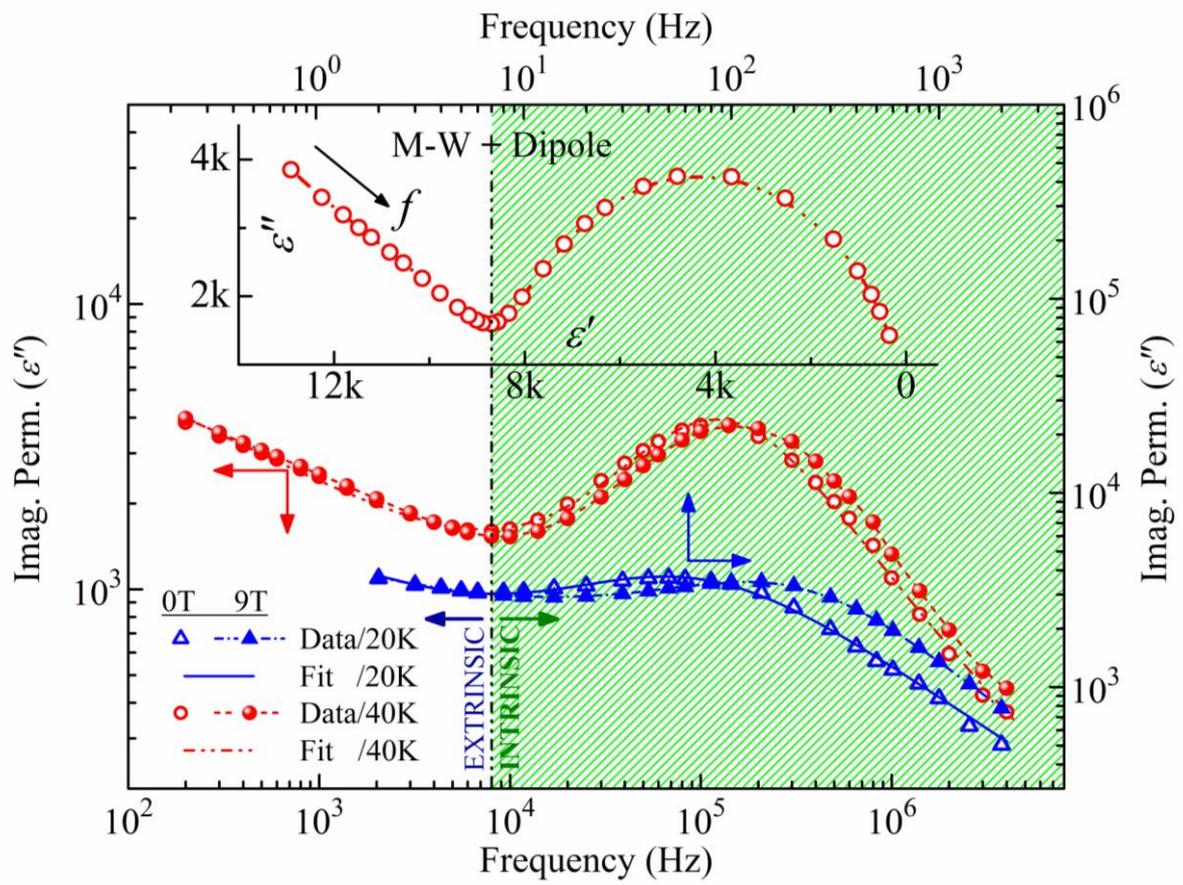

**Fig.6**

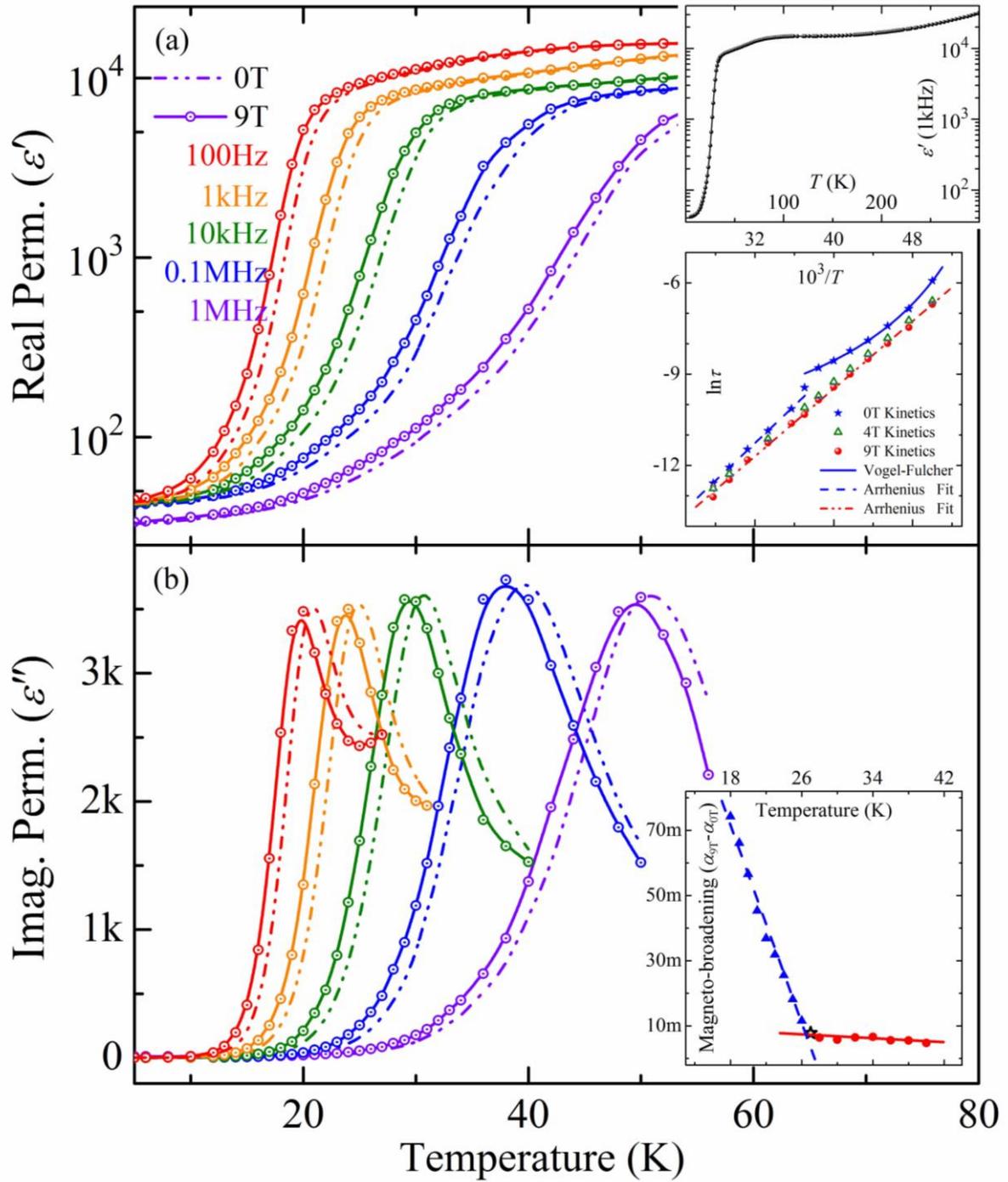

**Fig.7**

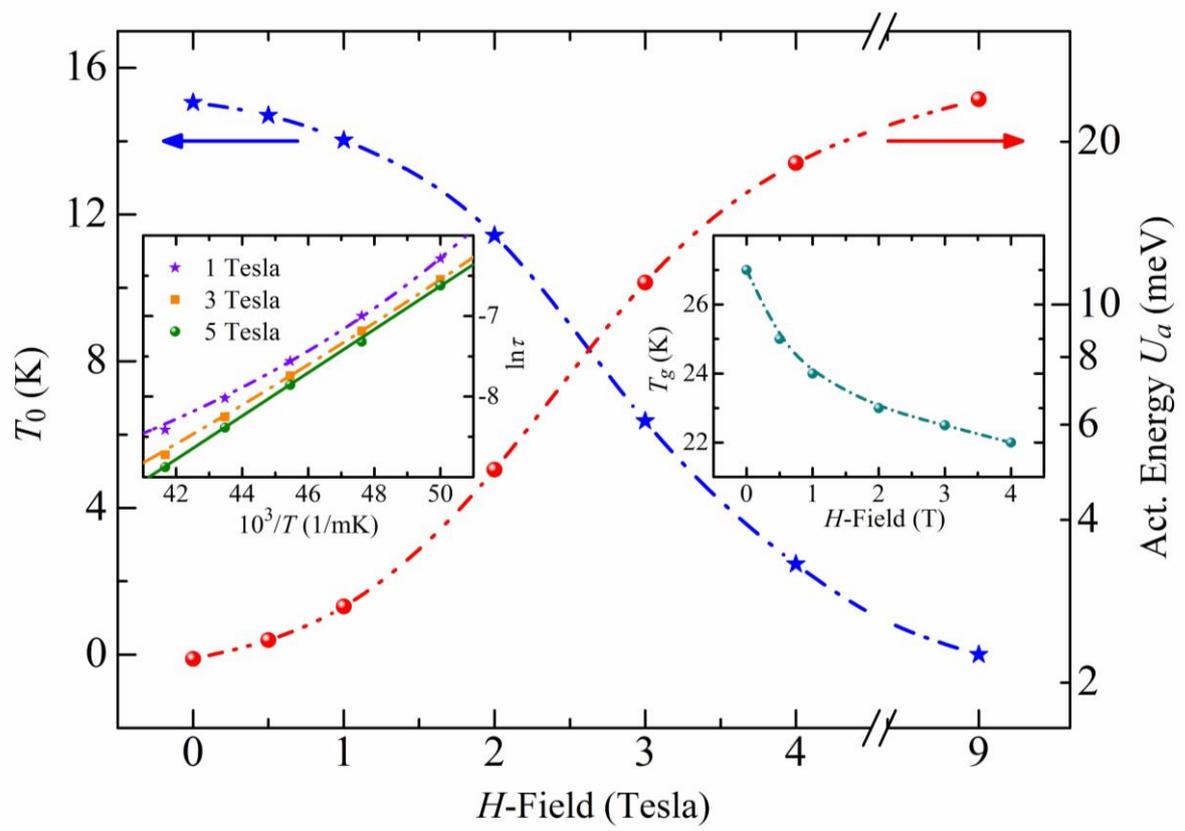

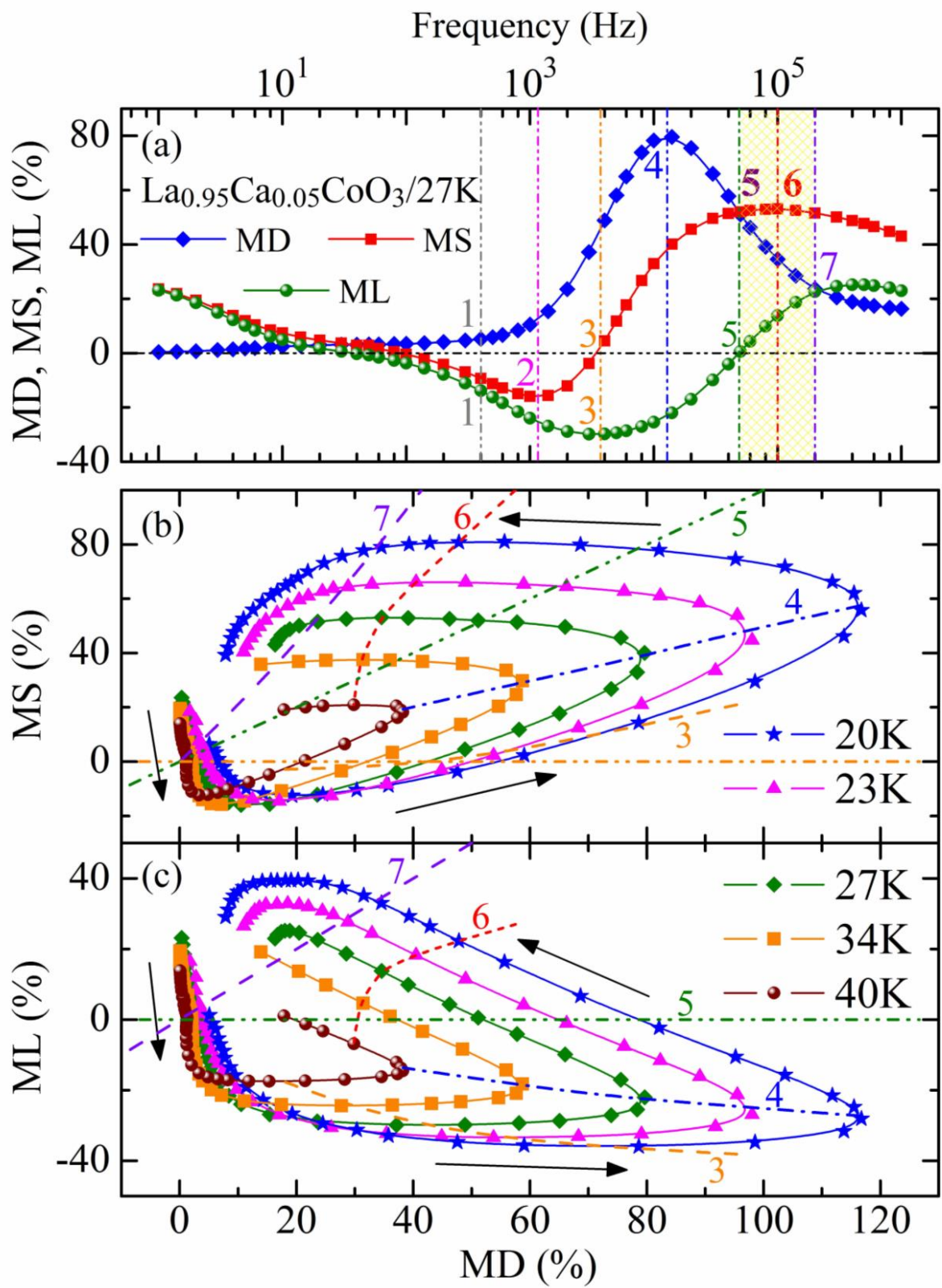

**Fig.9**

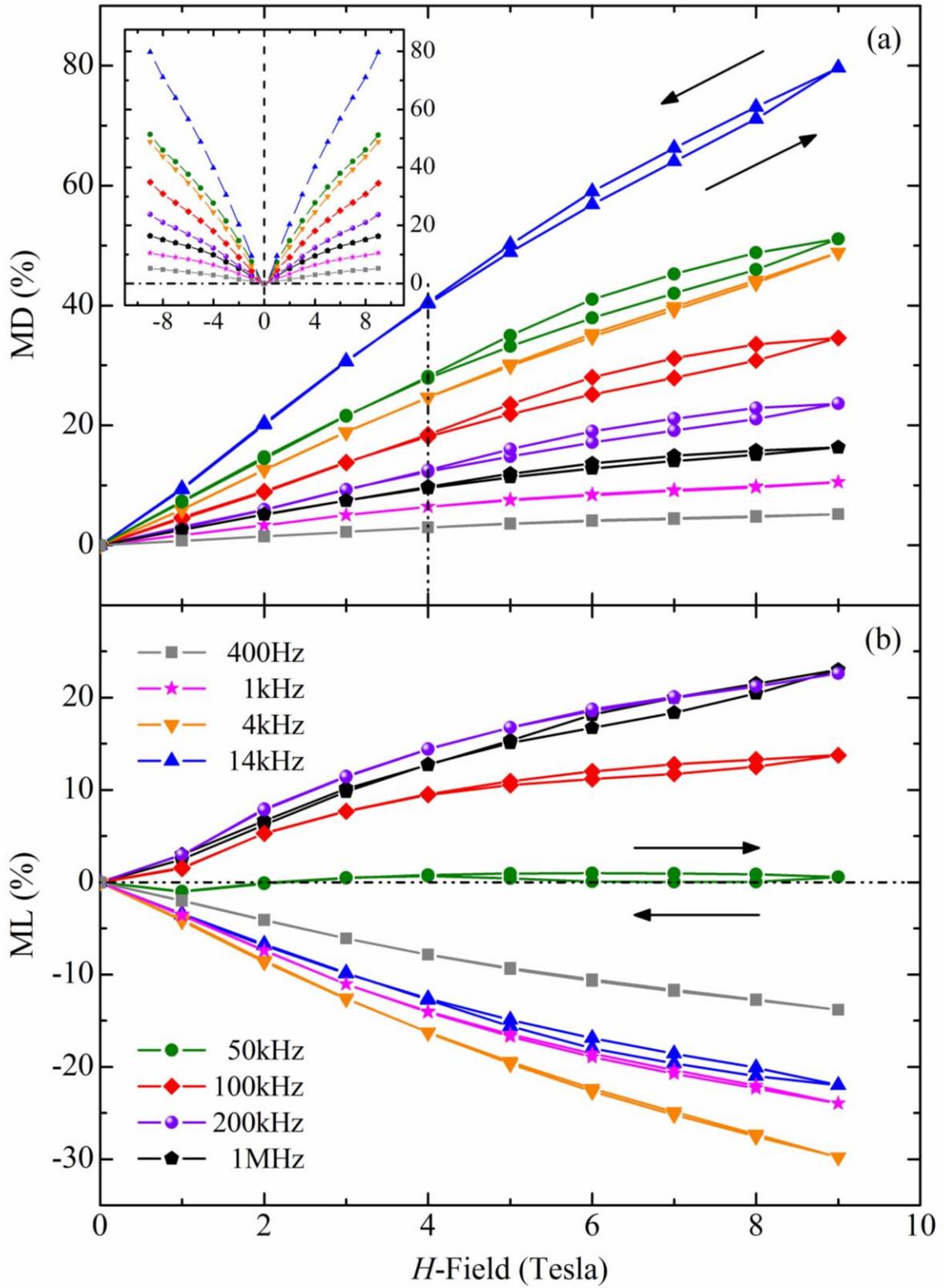

**Fig.10**

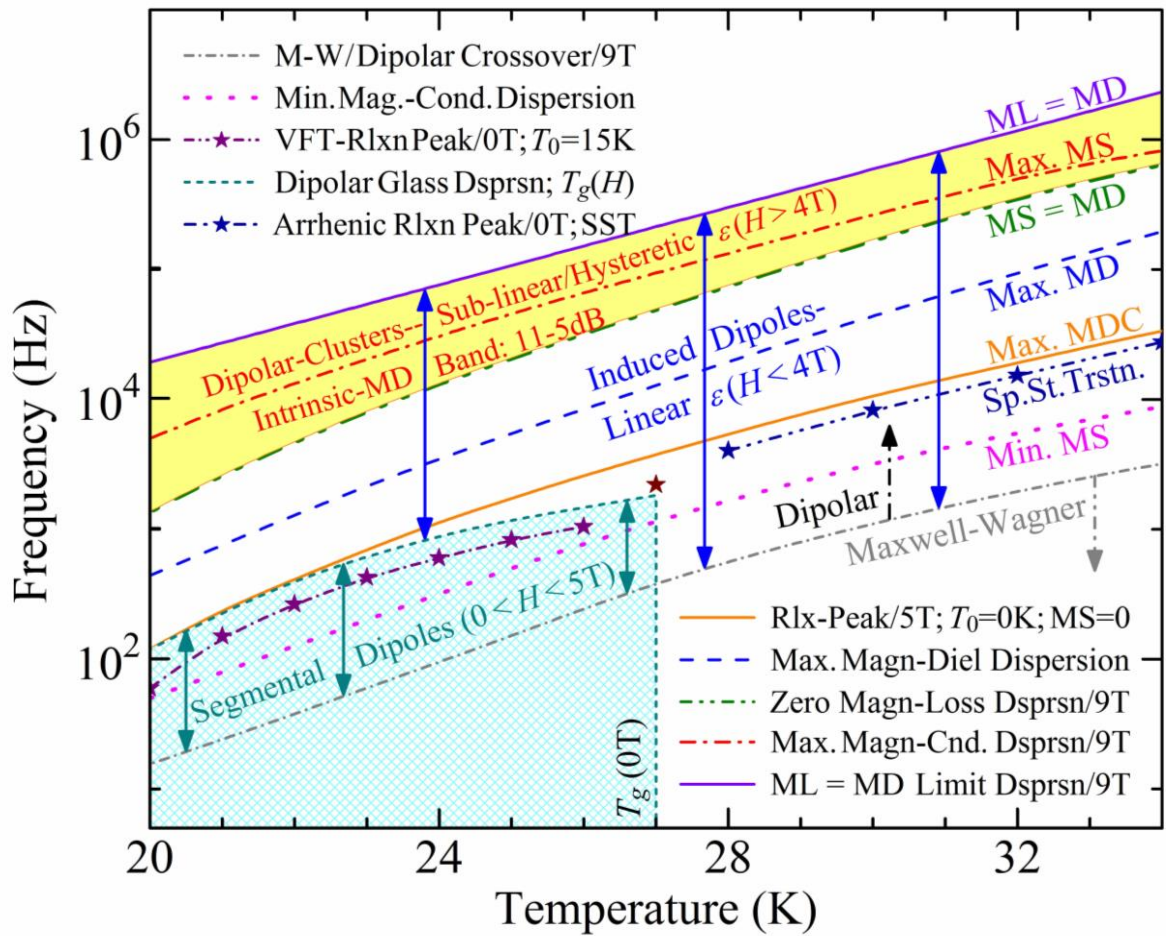